\documentclass{aa}
\usepackage{graphicx}
\usepackage{txfonts}
\begin{document}
   \title{Study of the Surface of 2003 EL$_{61}$: the largest carbon-depleted object in the trans-neptunian belt}

   \subtitle{}
\author{N. Pinilla-Alonso
          \inst{1}
	  \and
	  R. Brunetto
	  \inst{2,3}
	  \and
	  J. Licandro
	  \inst{4}
	  \and
	  R. Gil-Hutton
	  \inst{5}
	  \and
	  T. L. Roush
	  \inst{6}
	  \and
	  G. Strazzulla
	  \inst{3}
	  }

   \offprints{N. Pinilla-Alonso}

  \institute{Fundaci\'on Galileo Galilei \& Telescopio Nazionale Galileo, P.O.Box 565, E-38700, S/C de La Palma, Tenerife, Spain.\\
              \email{npinilla@tng.iac.es}
         	\and
	     Institut d'Astrophysique Spatiale, Universit\'e Paris-Sud, b\^atiment 121, 91405 Orsay Cedex, France.\\
 		\and
             INAF-Osservatorio Astrofisico di Catania, Via S. Sofia 78, I-95123, Catania, Italy.
		\and
	     Instituto de Astrof\'{\i}sica de Canarias, c/V\'{\i}a L\'actea s/n, E38205, La Laguna, Tenerife, Spain.\\
         	\and
	     Complejo Astron\'omico El Leoncito (Casleo) and San Juan National University, Av. Espa\~na 1512 sur,J5402DSP, San Juan, Argentina
         	\and
	     NASA Ames Research Center, MS 245-3, Moffett Field, CA 94035-1000, USA\\
}
   \date{Received ; accepted }

% \abstract{}{}{}{}{}
% 5 {} token are mandatory

  \abstract
  % context fh
   {2003 EL$_{61}$ is the largest member of a group of trans-neptunian objects (TNOs) with similar orbits and 'unique' spectral characterisitcs: neutral slope in the visible and the deepest water ice absorption bands ever observed in the trans-neptunian belt (TNb). Studying the surface of 2003 EL$_{61}$ provides useful constrains on the origin of this particular group of TNOs and on the outer Solar System's history.}
    {To study the composition of the surface of 2003 EL$_{61}$.}
    {We present visible and near-infrared spectra of 2003 EL$_{61}$ obtained with the 4.2m WHT and the 3.6m TNG telescopes at the Roque de los  Muchachos Observatory (Canary Islands, Spain). Near infrared spectra were obtained at different rotational phases covering almost one complete rotational period. Spectra are fitted using scattering models based on Hapke theory and constraints on the surface composition are derived.}
    {The observations confirm previous results: 2003 EL$_{61}$ spectrum is neutral in color and presents very deep water ice absorption bands. Meanwhile, they provide us with some new facts about the surface of this object, no significant variations in the spectral slope (in the near-infrared) and in the depth of the water ice absorption bands at different rotational phases are evident within the S/N, suggesting that the surface of 2003 EL$_{61}$ is homogeneous. The scattering models show that a 1:1 intimate mixture of crystalline and amorphous water ice is the most probable composition for the surface of this big TNO, and constrain the presence of other minor constituents to a maximum of 8\%\ }
    {The derived composition suggests that: a) cryovolcanism is unlikely to be the resurfacing process that keeps the surface of this TNO, and the other members of this population, covered mainly by water ice; b) the surface is older than 10$^{8}$ yr. This constrains the time scale of any catastrophic event, like the collision suggested to be the origin of this population, to at least 10$^{8}$ yr; c) the surface of 2003 EL$_{61}$ is depleted of carbon bearing species. According to the orbital parameters of the population, this makes it a possible source of carbon-depleted Jupiter family comets.}


\maketitle
%
%________________________________________________________________

\section{Introduction}

2003 EL$_{61}$ is the fourth largest known TNO and one of the most peculiar ones. It is the largest member of a group of TNOs with surfaces composed of almost pure water ice and moving around the Sun in a narrow region of the orbital parameters space (41.6 $<$a$<$ 43.6 AU, 25.8 $<$i$<$ 28.2 deg., 0.10 $<$e$<$ 0.19) (Pinilla-Alonso et al., \cite{Pinilla-AlonsoRR43}). Brown et al. (\cite{Brownfamily}) claim that this group is a family of fragments produced by a giant collision that happened in the trans-neptunian belt (TNb). The spectra of all these objects show the same characteristics, they are dominated by neutral slopes in the visible and large water ice absorption bands in the near-infrared (the deepest ever observed in the TNb, Pinilla-Alonso et al., \cite{Pinilla-AlonsoRR43}).

The study of the surface composition of these TNOs could provide important clues to understand the origin of this group, the role of
collisions in sculpting the TNb,  and the resurfacing processes that take place in this region of the Solar System. In particular, since 2003 EL$_{61}$ is the brightest one, it is the best candidate for such a study. Some works have been done with this purpose since the discovery of this object.

Rabinowitz et al., (\cite{RabinEL61}) show a photometric study of  2003 EL$_{61}$ revealing that this body has an unambiguous double-peak light curve. They discuss two different explanations for this curve, assuming this body is an homogeneous elongated ellipsoid whose shape determines the light curve variations and assuming it is a flattened spheroid where light variations are related with albedo patterns on its surface.

Tegler et al. (\cite{TegEL61}) obtain a  visible spectrum of 2003 EL$_{61}$ and show that it is featureless, with a neutral slope, they also report the possible existence of a band at 5773 \AA {} probably due to O$_{2}$-ice, anyway this feature has not yet been confirmed as its confirmation requires a higher signal precision of the contiunum in the spectrum.

Trujillo et al. (\cite{TrujEL61}) present a near-infrared spectrum of 2003 EL$_{61}$ and using  Hapke models (Hapke \cite{HapkeModels}) they claim that it can be modeled considering crystalline water ice and a blue component neccesary to fit the spectrum above 2.35$\mu$m like HCN or hydrated tholins. Merlin et al. (\cite{MerEL61}) present a new visible and near-infrared spectrum of 2003 EL$_{61}$ and fit them with Skhuratov models that suggest that the surface of the TNO is essentially composed of water ice, mainly in crystalline state, but in a small fraction of the surface the water ice could be amorphous.

In this paper we present new spectroscopic observations from 2003 EL$_{61}$ in the visible and near-infrared (Section 2). In Section
3, we present a collection of near-infrared spectra of 2003 EL$_{61}$ taken at different rotational phases and study the homogeneity of its surface. In Section 4 we model the spectrum of 2003 EL$_{61}$ and detail the physical implications for the object composition. We discuss previous results in Section 5 and summarize the conclusions in Section 6.

%__________________________________________________________________
\section{Observations}

We observed 2003 EL$_{61}$ on 2005 August 1.92 UT simultaneously with two telescopes at the ``Roque de los Muchachos Observatory" (ORM, Canary Islands, Spain), namely the 4.2m William Herschel (WHT) and the Italian 3.6m Telescopio Nazionale Galileo (TNG), and only in the near-infrared with the TNG on 2006 February 25, from 01:21 to 04:33 UT. Both nights were photometric.

\subsection{Visible Spectrum}

The visible spectrum (0.35-0.98 $\mu$ m) was obtained using the low resolution gratings (R300B and R158R with dispersions of 0.86 and 1.63 \AA/pixel, respectively) of the double-armed spectrograph ISIS at WHT, and a 5" slit width oriented at the parallactic angle (the position angle for which the slit is perpendicular to the horizon) to minimize problems with differential atmospheric refraction. The tracking was at the TNO proper motion. Three 300s spectra were obtained by shifting the object by 10" in the slit to better correct the fringing. Calibration and extraction of the spectra were done using IRAF and following  standard procedures (Massey et al., \cite{Massey1992}).

The three spectra of the TNO were averaged. Spectra of the solar-analogue star BS4486 and the G2 Landolt (SA) 107+998 (Landlot, \cite{Landolt}) were obtained the same night at a similar airmass just before and after the TNO spectrum, and used to obtain the final reflectance spectrum of the TNO, normalized at 0.5$\mu$m shown in Fig.\ref{Fig1}.

\subsection{Infrared Spectrum}

The near-infrared spectra were obtained using the high throughput, low resolution spectroscopic mode of the Near-Infrared Camera and Spectrometer (NICS) at the TNG,  with an Amici prism disperser. This mode yields a complete 0.8-2.5$\mu$m spectrum. We used a 1.5"  and a 0.75" wide slit in 2005 and 2006 observations respectively, corresponding to a spectral resolving  power R$\sim$34 and R$\sim$70 along the spectrum. The identification of the TNO was done by taking a series of images through the J$_s$ filter ($\lambda_{cent}$=1.25$\mu$m). 2003 EL$_{61}$ was identified as a moving object at the predicted position and with the predicted
proper motion. The slit was oriented at the parallactic angle and the tracking was at the TNO proper motion. We used the observing and reduction procedure described by Licandro et al. (\cite{Licreduc}).

The acquisition consisted of a series of images (3 in 2005 and 1 in 2006 observations respectively) of 90 seconds exposure time in one slit position (position {\em A}) and then offsetting the telescope by $10"$ in the direction of the slit (position {\em B}). This process was repeated obtaining several {\em AB} cycles up to a total exposure time of 1080 and 5220 seconds in 2005 and 2006, respectively.

To correct for telluric absorption and to obtain the relative reflectance, the G stars Landolt (SA) 98-978, Landolt (SA) 102-1081, Landolt (SA) 107-998 ( Landlot, \cite{Landolt}) were observed at different airmasses along the night, before and after the TNO observations, and used as  solar analogue stars.

\begin{figure}
	\centering
	\includegraphics[width=\columnwidth]{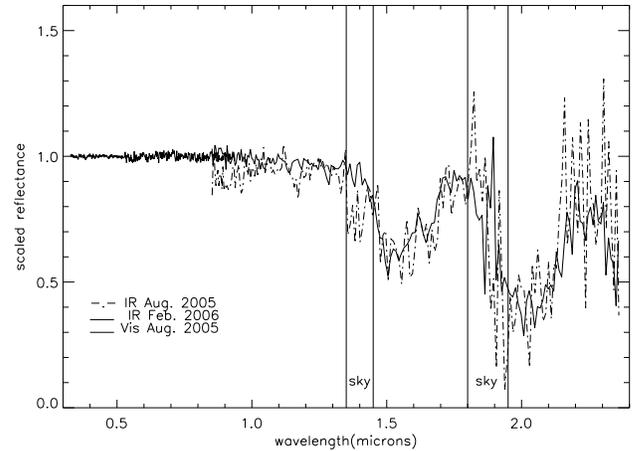}
	\caption{VIS (2005) and NIR (2005 and 2006) spectra of 2003 EL$_{61}$. IR spectra match within the errors. Red slope bellow 1 micron in 2005 NIR spectrum can be a result of the centering of the object in the slit. Uncertainties are given by the point to point variation.}
 	\label{Fig1}
 \end{figure}

The spectra of 2003 EL$_{61}$ were divided by the spectra of the solar analogue stars, and then normalized to unity around 1.0$\mu$m thus obtaining the scaled reflectance. In Fig.\ref{Fig1} The resulting spectra obtained by combining all the extracted $AB$ of each night and normalized to join the visible spectrum around 0.9$\mu$m are presented.

Around the two large telluric water band absorptions the S/N of the spectrum is very low. Even in a rather stable atmosphere, the telluric absorption can vary between the object and solar analogue observations introducing false features. Therefore, any spectral structure in the 1.35-1.46 and 1.82-1.96$\mu$m regions can produce false features.  There are also a few more less deeper telluric absorption regions: 0.93-0.97, 1.10-1.16, 1.99-2.02, and 2.05-2.07$\mu$m. Features in these regions must be carefully checked.

% FIGURA CON EL VISIBLE + EL IR DE 2005 + EL IR DE 2006

\begin{table}
\centering
\begin{tabular}{l c c c c}
\hline
\hline

% Spectrum   Fecha y hora Texp	Airmass	FaseRotacion

Spectrum & Date & texp(s) & airmass & rot. phase\\ \hline\hline
Aug. 2005 &  1.92-1.93 & 1080 & 1.90 & ?\\ \hline
2006 phase1 & 25.06-25.07 & 1080 & 1.30 & 0 - 0.10\\ \hline %25.06 - 25.071.35-1.73
2006 phase2 & 25.08-25.10 & 1080 & 1.19 & 0.13 - 0.32\\ \hline %1.87-2.27
2006 phase3 & 25.13-25.15 & 720 & 1.05 & 0.47 - 0.55\\ \hline %3.18-3.48
2006 phase4 & 25.15-25.17 & 540 & 1.02 & 0.59 - 0.73\\ \hline %3.67-4.18
2006 phase5 & 25.18-25.19 & 720 & 1.01 & 0.77 - 0.84\\ \hline%4.37-4.62

\end{tabular}
 \caption{The date, total exposure time, mean airmass, and rotational phase of the near infrared spectrum obtained in 2005 and the five combined spectra from the series obtained on 2006 February 25 (see text).}
\label{Table1}
\end{table}

\section{Analysis of the spectrum}

The spectrum of 2003 EL$_{61}$ shows the characteristics described in previous papers (Merlin et al., \cite{MerEL61} and references therein) (a) The visible is featureless within the S/N and neutral (the spectral slope, computed between 0.53 and 1.00$\mu$m, is $S_{vis}'= 0.0\pm 2\%/1000$\AA), such a neutral slope together with the high albedo estimated for this TNO ($\rho_{v}$ = 0.6-0.8, Rabinowitz et al., \cite{RabinEL61}) is indicative of the lack of complex organics (typically red in color produced by the irradiation of hydrocarbons and/or alcohols) on its surface; (b) The slope of the continuum is blue in the near-infrared; (c) It presents two deep absorption bands centered at 1.52 and 2.02$\mu$m, indicative of water ice; (d) The band at 1.65$\mu$m, typical of crystalline water ice, appears in all our spectra in the near-infrared; (e) There is a clear absorption above 2.2$\mu$m.; (f) There is no absorption band in the 0.6-0.8$\mu$m spectral region indicative of phyllosilicates (Fornasier et al., \cite{Fornetal2004}).

Due to the S/N of our spectrum in the visible we cannot confirm the presence of a feature at 0.58$\mu$m that has been suggested by Tegler et al., (\cite{TegEL61}) and would be related to the presence of O$_{2}$-ice in the surface.

\subsection{Rotational variation}

\begin{figure}
	\centering
	\includegraphics[width=\columnwidth]{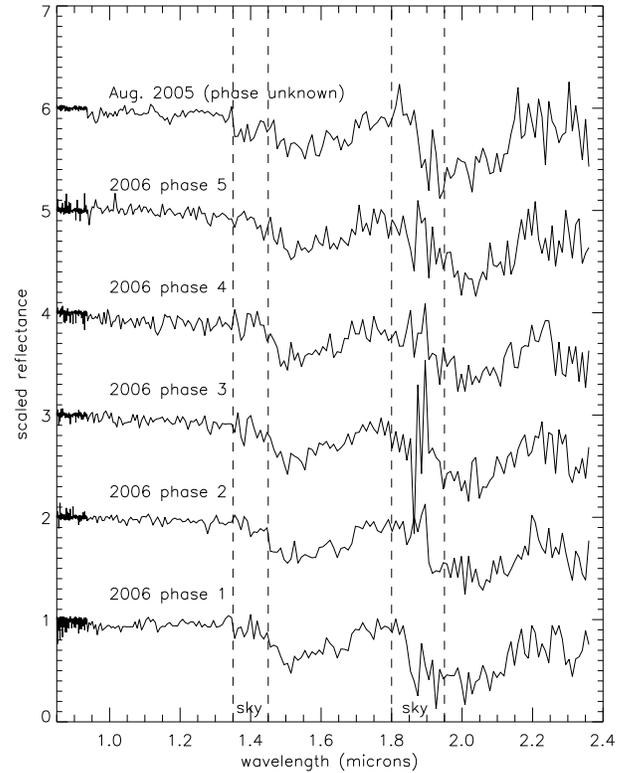}
	\caption{Five spectra obtained on 2006 February 25 covering different rotational phases (see text for details) and the spectrum obtained in 2005. Uncertainties are given by the point to point variation. Notice that all look very similar suggesting that the surface is homogeneous within our errors (see text for more details).}
 	\label{Fig2}
\end{figure}

On 2006 February 25, we obtained a series of near-infrared spectra of 2003 EL$_{61}$ from 01:21 to 04:33 UT. The rotation period of this TNO is 3.9 hrs. (Rabinowitz et al., \cite{RabinEL61}), so our observations cover about  80 \% of the rotation of 2003 EL$_{61}$ and allow us to study possible surface inhomogeneities. To do that we combined five sub-samples of this series to produce five spectra covering different rotational phases (see Table \ref{Table1} for details). Fig.\ref{Fig2} shows these five spectra and the spectrum obtained in 2005.

\begin{figure}
	\centering
	\includegraphics[width=\columnwidth]{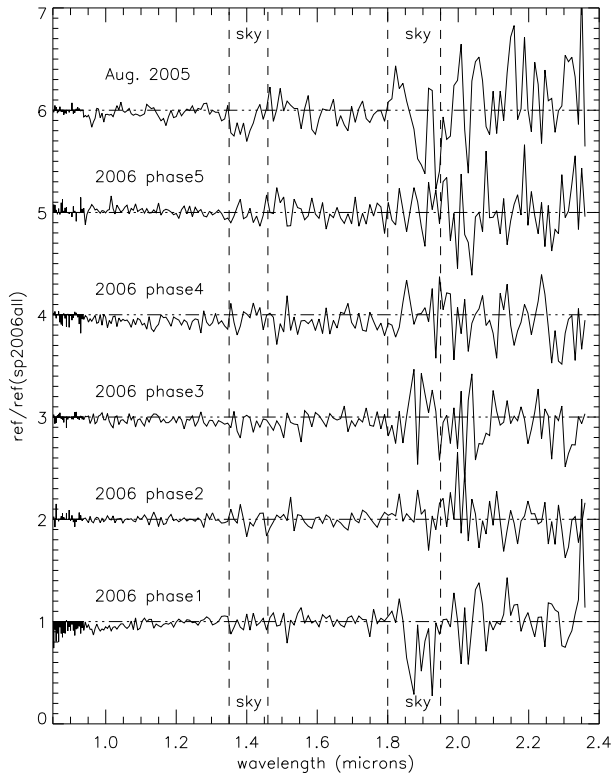}
		\caption{Ratio between the spectra shown in Fig.\ref{Fig2} and the \textit{sp2006all} spectrum. Uncertainties are given by the point ot point variation.Notice that there is no systematic deviation within the limit of the noise for none of the spectra obtained at different phases.}
 	\label{Fig3}
 \end{figure}

There is no evidence of large inhomogeneities on the surface of the TNO. Every single spectrum shows the same features within the S/N. Fig.\ref{Fig3} shows the ratio between these spectra and the spectrum of 2003 EL$_{61}$ represented in Fig.\ref{Fig1} (labelled \textit{sp2006all}). Notice that there is no systematic deviation within the limit of the noise for none of the spectra obtained at different phases, neither in the slope nor in the spectral bands.

Finally, to quantify any possible variation in the slope and/or band depths of the spectra we computed two parameters:

\begin{enumerate}
\renewcommand{\labelenumi}{(\alph{enumi})}
 \item The normalized infrared reflectivity gradient $S_{ir}'[\%/1000$ \AA$]$ over the 0.85-1.3$\mu$m range. To compute it we fit the continuum between 0.85 and 1.3$\mu$m with a straight line. This fit is sensitive to the first and last points of the spectrum but the error introduced by this factor is negligible compared to the systematic errors.
 \item The depth of the bands, \textit{D}, with respect to the continuum of the spectrum (fitted in the infrared using a cubic spline) as: \textit{D=1-R$_{b}$/R$_{c}$}, where \textit{R$_{b}$} is the reflectance in the center of the band (1.52 or 2.02$\mu$m) and \textit{R$_{c}$} is the reflectance of the continuum in the same points (1.52 or 2.02$\mu$m, respectively).
\end{enumerate}

Table \ref{Table2} presents the values of these parameters obtained from each spectrum. The mean $S_{ir}'$=-0.9$\pm$1.8. The largest deviation from this mean value corresponds to the spectrum obtained at phase 1 and it is $2.5\%/1000$\AA. Considering that, due to systematic errors (e.g. centering uncertainties of the TNO and/or solar analogues in the slit...) uncertainties of $2\%/1000$\AA {} in the slope are usual, we should conclude that the spectral slope is similar at all rotational phases. Lacerda et al. (\cite{LacerdaEL61}) found a small variation of 0.035 magnitudes in the B-R color of 2003 EL$_{61}$ correlated with the rotation of the TNO. They suggest that this is produced by a region of the surface with lower albedo and redder in color. This change in color corresponds to a reddening of $S'$ of about 1-2$\%/1000$\AA {} which is smaller than the uncertainties of our measurements so we cannot confirm Lacerda et al. results from the study of the spectra.

In the same way, the mean depth of the water absorption bands \textit{D}$_{1.52}$ and \textit{D}$_{2.02}$ are 33.7$\pm$2.5 and 53.5$\pm$4.0 respectively. The errors of the measured band depths for each spectrum are larger than the deviation from the mean, so we can conclude that there are no variations of the depths of the water absorption bands larger than \%.
We conclude that the surface of 2003 EL$_{61}$ is homogeneous in a large scale. This reinforces the hypothesis that its shape is like a very elongated ellipsoid.

\begin{table}
\centering
\begin{tabular}{l c c c c c}
\hline
\hline

name & $S_{ir}'$ & \textit{D}$_{1.52}$ & \textit{$\varepsilon_{1.52}$} & \textit{D}$_{2.02}$ & \textit{$\varepsilon_{2.02}$} \\  \hline\hline

Aug. 2005 & 1.07 & 31.13 & 7.2 & 49.7 & 13.5 \\  \hline
sp 2006all & -1.40 & 34.0 & 4.8 & 53.8 & 6.6 \\  \hline
2006 phase1 & 1.61  & 33.9 & 6.7 & 55.8 & 10.0 \\  \hline
2006 phase2 & -1.77 & 33.3 & 5.6 & 51.3 & 10.2 \\  \hline
2006 phase3 & -2.12 & 38.3 & 7.0 & 54.6 & 10.6 \\  \hline
2006 phase4 & -2.32 & 32.4 & 9.2 & 49.5 & 13 \\  \hline
2006 phase5 & -1.97 & 33.1 & 7.8 & 59.8 & 10.6 \\  \hline

\end{tabular}
 \caption{ Normalized reflectivity gradient $S_{ir}'$ and depth of the 1.52 and 2.02 $\mu$m bands \textit{D},  as defined in the text, for the all the considered infrared spectra . Uncertainties of $2\%/1000$\AA {} in the slope are typical due to systematic errors (see text for details). $\varepsilon$ are the associated errors of \textit{D}.}

% Pendiente calculada entre 0.85 y 1.3 sólo el IR con el programa unirEL61mio.pro

\label{Table2}
\end{table}

\section{Spectral Models}

In this section, we use the Hapke scattering model (Hapke, \cite{HapkeModels}; Hapke, \cite{Hapke93}) to fit the spectrum of 2003 EL$_{61}$ and derive conclusions about its surface composition. In particular we test whether the spectrum is consistent with the presence of amorphous ice and investigate the presence of other minor components.

To fit the model we use our best S/N spectrum named \textit{sp2006all}, obtained by combining all the spectra acquired in the 2006 run as explained in Section 2. To increase the S/N, we merge our spectrum of 2003 EL$_{61}$ with that of Trujillo et al., \cite{TrujEL61}. Notice that there are no significant differences between our spectrum and the already published data. 

It is important to stress that the knowledge of the optical constants is often poor, and this affects the calculation and the model results. Thus, from visible and near-infrared observations it is usually impossible to retrieve a unique composition and the fits we show must be regarded as possible solutions.
 
Although the models are capable of incorporating multiple grain sizes for each component, such an effort is beyond the scope of the current report. Additionally, given the inherent uncertainties in the optical constants we consider relatively simple models consisting of a single grain size for each component adequate for our effort here.The models shown in this section are those with the smallest reduced chi square value.

For crystalline water ice we use optical constants from Grundy \& Schmitt, (\cite{Grundywater}) (obtained at T= 40 K) in the 1.0-2.5$\mu$m spectral region and from Warren, (\cite{Warren84}) (obtained at T$\approx$ 266 K) at shorter wavelengths. Unfortunately no data are available at appropriate temperatures in the visible region. The intensity of the 1.65 µm band that characterizes  crystalline ice depends on the temperature, being larger at lower T. The use of optical constants at 40 K is then necessary being such a T appropriate for TNOs. However it is important to outline that ice directly formed at 40 K is amorphous and does not exhibit the 1.65 µm band. Then the observation of the crystalline form implies that ice present on 2003 EL61 experienced higher temperatures (e.g. Baratta et al. \cite{Baratta91}).

For amorphous water ice, we use optical constants from Schmitt et al., (\cite{Schmitt}) (obtained at T= 38 K) in the near-infrared and those of crystalline water ice from Warren, (\cite{Warren84}) at wavelengths shorter than 1$\mu$m because, at our best knowledge, there are not available optical constants for amorphous ice in this spectral region.

We emphasize that the effects of ion irradiation produce an amorphous ice that is very similar to a deposited amorphous ice (Leto \& Baratta, \cite{Leto03}), and this allows us to use optical constants of deposited amorphous ice to mimic the amorphization induced by cosmic rays on TNOs. However, differences may be present, so future studies should focus on the derivation of optical constants of irradiated ice.

In this paper we also fit models considering the following minor components, CH$_{4}$, poly-HCN, NH$_{3}$, Ammonia Hydrate, Kaolinite and Ortho-Pyroxene. For them we use the optical constants from Grundy et al., (\cite{GrundyMet}), Khare et al., (\cite{Khare}), Brown et al., (\cite{BrownHA}), Sill et al., (\cite{Sill}),Clark et al., (\cite{Clark}) and Brunetto et al. (\cite{Brunetto-Pyroxenes}), respectively.

\subsection{ Crystalline and amorphous water ice spatial mixtures}

\begin{figure}
	\centering
	\includegraphics[width=\columnwidth]{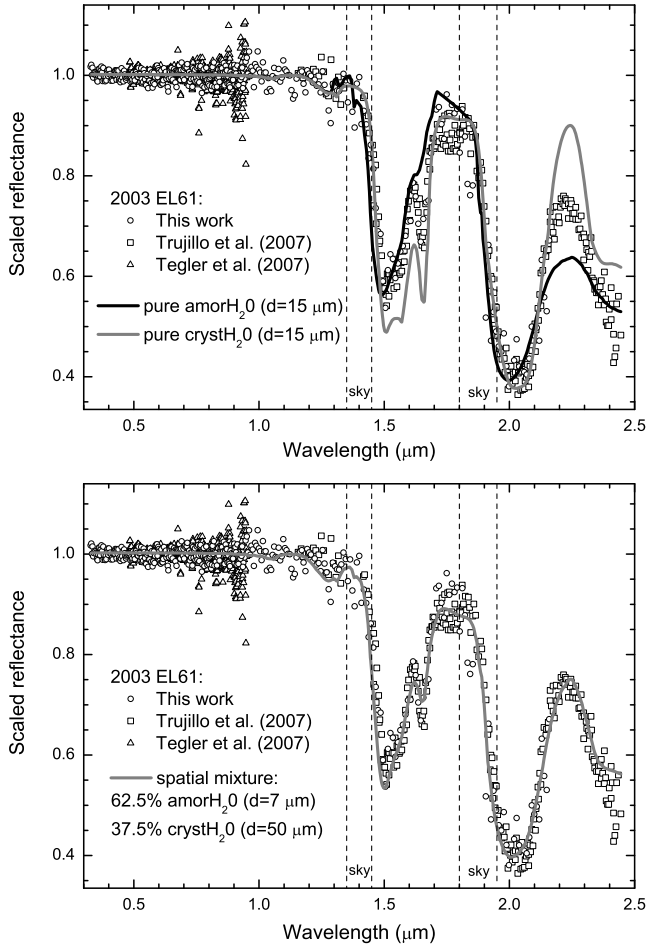}
	\caption{Upper panel: the spectrum of 2003 EL$_{61}$ compared with two synthetic spectra, using only crystalline or only amorphous water ice. Lower panel: the spectrum of 2003 EL$_{61}$ fitted by a spatial mixture of crystalline and amorphous ice; either trials, the fit is not satisfactory.}
 	\label{Fig4}
 \end{figure}

We first tried to fit the spectrum with pure crystalline water ice (Fig.\ref{Fig4}. upper panel), the fit fails mainly at reproducing the 1.52$\mu$m and 2.02$\mu$m bands at the same time and at reproducing the depth of the 2.2$\mu$m band. We investigate the possibility of reproducing the spectrum with only pure amorphous water ice (Fig.\ref{Fig4}. upper panel). Amorphous water ice is bluer at the longer wavelengths ($\geq 2.2\mu$m) than crystalline water ice but it does not reproduce the spectrum well either. This is chiefly because the 1.65$\mu$m feature is absent in amorphous ice and the 1.5 and 2.0$\mu$m are at too short a wavelength compared to the observational data.

Then, we investigated a model of a mixture of amorphous and crystalline ice. A first approach is to assume that crystalline and amorphous ice are spatially segregated on the surface. Thus, we applied a spatial mixture. The results are shown in the lower panel of Fig.\ref{Fig4}. The best fit is a model of two segregated components: 37.5\%\ of crystalline water ice (with grain size of 7$\mu$m) and 62.5\%\ of amorphous water ice (with grain size of 50$\mu$m).
In the visible range the mixture produces a flat spectrum and a geometric albedo of about 0.7, which is in agreement with previous results.

The fit curve is not satisfactory because, even if several features of the spectrum are well reproduced it fails to reproduce the depth of the 1.65$\mu$m band.\\

\subsection{Crystalline and amorphous water ice intimate mixtures}

\begin{figure}
	\centering
	\includegraphics[width=\columnwidth]{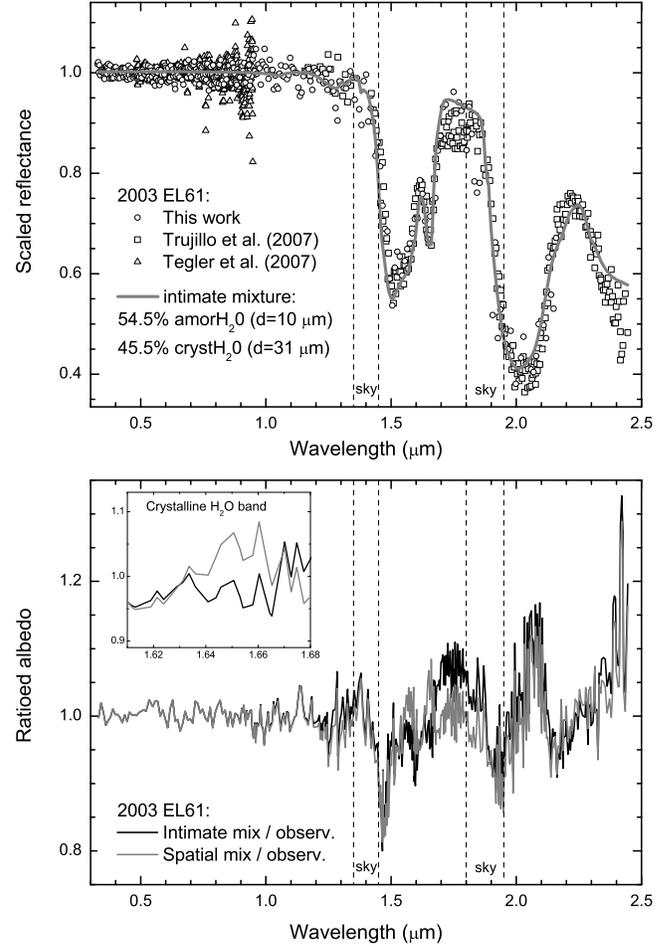}
	\caption{Upper panel: the spectrum of 2003 EL$_{61}$ fitted by an intimate mixture of crystalline and amorphous ice. Lower panel: plot of the ratio between the spatial and intimate mixture models, and the observed spectrum; the regions where discrepancies appear are evident. The insert shows the region of the 1.65$\mu$m feature, where an intimate mixture is clearlly better.}
 	\label{Fig5}
 \end{figure}

Since the spatial mixture was not satisfactory, we tested an intimate mixture of amorphous and crystalline ice. The results are shown in the upper panel of Fig.\ref{Fig5}; the model corresponds to a 54.5\%\ of amorphous ice (grain size of 10$\mu$m) and a 45.5\%\ of crystalline ice (grain size of 31$\mu$m). This model produces slightly different grain sizes for amorphous and crystalline ice but considering the errors in the optical constants  the differences in size may not be significant.
In the visible range, our model produces a flat spectrum and a geometric albedo of about 0.7.

This fit reproduces the data better than the spatial mixture in particular the 1.65$\mu$m absorption band (Fig. \ref{Fig5}, lower panel), but it reduces the quality of the fit in the 1.7-1.85$\mu$m range. However, the uncertainties of the optical constants in the 1.7-1.85$\mu$m range are larger than in the 1.65$\mu$m spectral band as the transmittance of the water ice is almost zero in this region. Some discrepancies are still present in the 2.2$\mu$m shoulder and in the 1.45$\mu$m and 2.4$\mu$m region. These discrepancies appear systematically in all our models and can be attributed to the uncertainties of the optical constants and observations.

Despite these discrepancies, this model reproduces better the most important features of the spectrum. Notice that an intimate mixture of crystalline and amorphous would imply a rather homogeneous surface, in agreement with our results based upon the study of the rotational variation.

A model that considers an intimate mixture of amorphous and crystalline water ice was also proposed by Merlin et al., (\cite{MerEL61}). Comparing our fit with that of Merlin et al., we find a much higher percentage of amorphous ice, which would indicate a longer exposure to ion irradiation, i.e. an older surface. Merlin et al. used very different grain size for amorphous and crystalline ice (more than a factor of 10), which is very hard to be explained, while in our model the difference is reduced to a factor of 3. Also they used a contribution from an amorphous carbon that is not included in our model. Considering that an small amount of amorphous carbon (3\%) would reduce considerably the geometric albedo, and that the albedo we derived from the model is $p_V$=0.7 consistent with the observations (Rabinowitz et al. \cite{RabinEL61}) we constrain the ammount of amorphous carbon to a very small fraction less than a few percent. 

We conclude we achieved a good fit of the spectrum of 2003 EL$_{61}$ that has stronger physical meaning, with an amorphous/crystalline ratio of 1:1 that indicates a competition between different physical processes on the surface.\\

\subsection{Multi-layer models}

Next step was to test a multi-layer system. This is an option that Hapke model offers, although not yet deeply explored(Brunetto \& Roush, \cite{Brunetto-Roush}). It consists on considering that from the moment that incident sun light reaches the surface of the body to the moment it leaves and is reflected to Earth, it travels trought layers composed of different materials characterized by different physical properties. Here we model a simple layered configuration using equations 7.45c, 8.89, 9.14, and 11.24 from Hapke (\cite{Hapke93}): the model calculates the reflectance from a system where a layer overlies an infinitely thick substrate.

This physical model arises from the consideration that 2003 EL$_{61}$ is homogeneously irradiated by solar wind and cosmic ions. Given tis situation, it is reasonable to imagine that different layers close to the surface have suffered a dose of irradiation that decreases with increasing depth causing the amorphization of water ice.

\begin{figure}
	\vspace{-85.pt}
        \centering
        \includegraphics[width=\columnwidth]{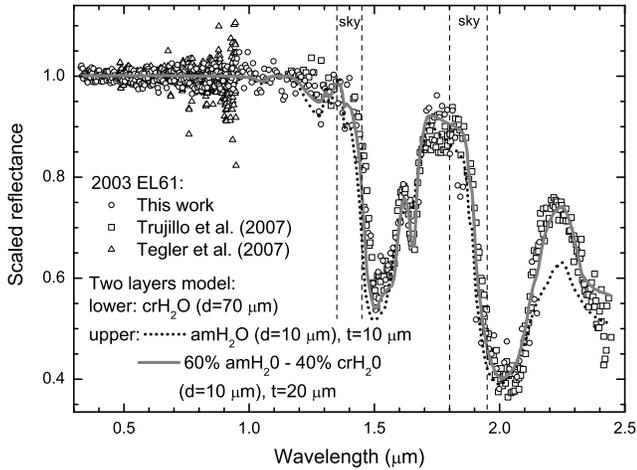}
	\vspace{-110.pt}
        \caption{Multi-layer models for 2003 EL$_{61}$: amorphous ice on top of
crystalline (dot) is unsatisfactory, while an intimate mixture of
amorphous and crystalline ice on top of crystalline (solid) is more
appropriate. Upper layer thickness and grain sizes are given for each
model.}
         \label{Fig6}
 \end{figure}

We initially considered a configuration where an amorphous mantle is on top of a crystalline substrate (Fig.\ref{Fig6}. dotted curve), but the result was unsatisfactory as using an upper layer of only 10$\mu$m amorphous ice is enough to mask the 1.65$\mu$m band of the underlying crystalline ice, in contrast to the observations of 2003 EL$_{61}$.

We then considered a configuration where the upper layer is composed of an intimate mixture of amorphous and crystalline water ice (Fig.\ref{Fig6}. solid curve). This model consits of an intimate mixture of 60\% amorphous and 40\% crystalline ice on top of an infinite crystalline ice substrate.

Although there are some remaining differences between the model and the spectrum this second model is closer to the spectrum
of 2003 EL$_{61}$. Actually, we find that all three major models (spatial mixture, intimate mixture, two-layers model) produce reduced chi squared of about 3. This would not allow to discriminate among them. Nevertheless, if we restrict to the region of the 1.65$\mu$m feature, we find values of: 3 for the spatial mixture; 2.4 for the intimate mixture; and 1.8 for the two-layers model.

These simple tests indicate that multi-layer models can be a useful tool to interpret TNOs spectra. Future efforts should focus on the possibility to model several layers, since ion irradiation would affect the icy layers gradually, and a sudden transition from amorphous to crystalline is not expected.\\

\subsection{Minor components}

One of the aims of this paper is to put constraints on the presence of minor components on the surface of 2003 EL$_{61}$. In this step we investigated the contribution of some species, using them as minor components in an intimate mixtures with amorphous and crystalline water ice. We consequently estimated upper limits for these species, and results are included in Table \ref{Table3}.

\begin{figure}
	
	\centering
	\includegraphics[width=\columnwidth]{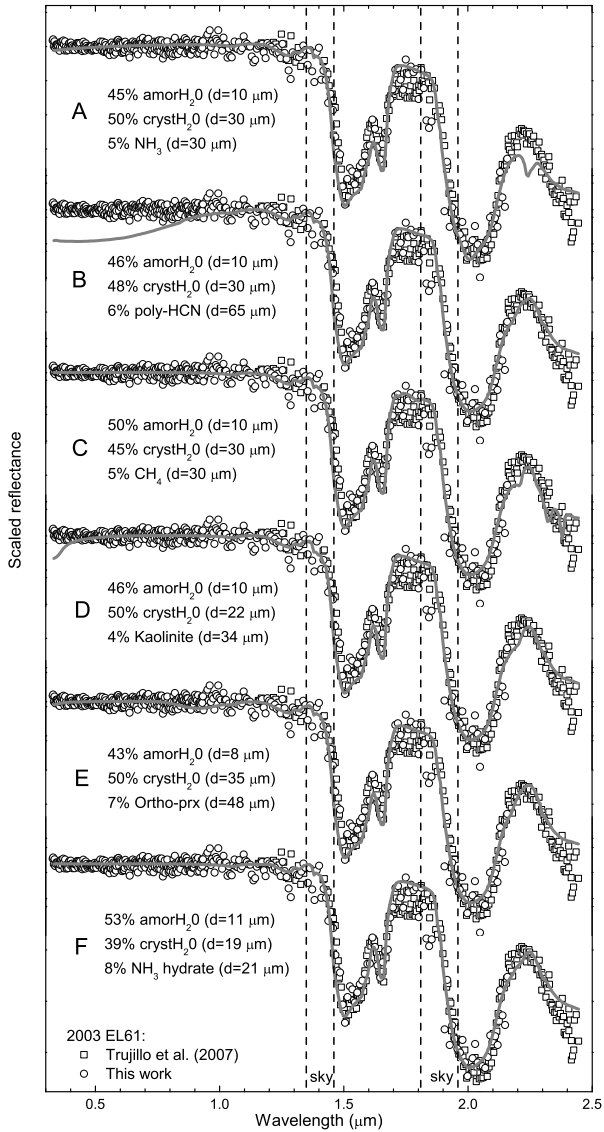}
	\caption{The effects of including minor components in the fit of 2003 EL$_{61}$ spectrum.}
 	\label{Fig7}
 \end{figure}

We put special attention to materials that have been proposed to be on the surface of this object. We found that none of the investigated components help to improve the fit quality (see Fig.\ref{Fig7}), furthermore, even small amounts of these materials change the shape of the model adding spurious features or reddening the slope, and impoverishing the quality of the fits.

The presence of ammonia and methane is discarded in a proportion larger than 5\% as we see from features introduced in the 2.2-2.3$\mu$m region (models A and C respectively).

The serpentine (not shown) and kaolinite contribution could help to produce a blue slope in the long wavelength region, but using percentages higher that 5-10\%\ introduces other features (especially in the visible, model D in Fig.\ref{Fig7}) that are not present in the observations. A simple 5\% contribution of poly-HCN (model B in Fig.\ref{Fig7}) is enough to produce non-neutral slopes in the visible range, which is in complete disagreement with the observations. Thus, the hypothesis of Trujillo et al., (\cite{TrujEL61}) of a strong contribution of these species has to be discarded. A large contribution of pyroxenes (model E in Fig.\ref{Fig7}) is discarded also as a small percentage of 7\% is enough to introduce an broad feature in the visible spectrum from 0.8 to 1.1$\mu$m.

In the case of hydrated ammonia (model F in Fig.\ref{Fig7}), the study is particularly interesting as it has been related to cryovolcanism. We find that a 8\% of hydrated ammonia impoverishes the quality of the fit as the reduced chi square value is 3.7, for the whole spectral range, and 2.3, if we restrict to the region around the 1.65$\mu$m, (far from the 3.0  and 1.8 values found for models with only water ice). In spite of this, adding hydrated ammonia is interesting when a feature around 2.2$\mu$m is detected (Cook et al. \cite{CookCharon}). Unfortunately, the relatively poor S/N of our spectrum does not allow us to clearly detect this feature, so we discard the presence of hydrated ammonia with an upper limit of 8\% but we condition this to the obtention of a better spectrum that confirms the presence or the absence of a feature around 2.2 microns.

%We want to empasize here that the presence of a species different from water ice than a particular percentage of hydrated ammonia implies is much smaller from the presence that a percentage of other material, as hydrated ammonia is form by ammonia and water ice in a proportion of 3:97.}

In conclusion, we discard the presence of other component on the surface but water to a high level of confidence. We stress that none of the components investigated helped us to improve the fit quality. In particular, the main discrepancies mentioned above remain. We think that discrepancies at 2.0 and 2.2$\mu$m can be attributed to uncertainties in the optical constants.\\

\begin{table}
\centering
\begin{tabular}{c c | c c}
\hline
\hline
Species & Upper limit  &  Species & Upper limit \\ \hline \hline
NH$_3$ &	5\%   & serpentine & 7\% \\  \hline
poly-HCN & 5\% & kaolinite & 4\% \\  \hline
CH$_4$ & 5\% & orthopyroxene & 7\% \\  \hline
CO$_2$ & 6\% & olivine & 5\% \\  \hline
NH$_4$OH & 8\% &\\  \hline
\end{tabular}
 \caption{Upper limits for possible minor components of the surface of 2003 EL$_{61}$.}
\label{Table3}
\end{table}

\section{Discussion}

In the previous Sections we find a surface composition for 2003 EL$_{61}$ of $>$92\% of pure water ice, and we discard the presence of a significant fraction of complex organics on its surface. In spite that this is not the first object showing a large fraction of water ice on its surface, TNOs with no traces of organics should not be common assuming that the original chemical composition of all TNOs is very similar i.e. objects composed of abundant water ice, some molecular ices like CO, CO$_2$, CH$_4$,N$_2$.

Laboratory experiments with astrophysical ice mixtures show that long term processing by high energy particles and solar radiation induces the loss of hydrogen and the formation of complex organic species, resulting in a dark and usually red mantle that covers the unprocessed original ices (Moore et al., \cite{Mooreetal83}; Johnson et al., \cite{Johnetal84}; Strazzulla \& Johnson, \cite{StrazzullaJo91}). In the case of a TNO, Gil-Hutton (\cite{Gil-Hutton}) shows that the time scale required to form this mantle of carbon residues, even when there is a competition between darkening by cosmic-rays and resurfacing due to impacts, is $\sim$ 10$^{9}$ years. If we consider this timescale for the development of a mantle, it is expected that a normal TNO begins to show the reddening of its spectrum, due to the presence of these organics species, in shorter periods.

In the particular case of 2003 EL$_{61}$ its spectrum does not show the expected presence of complex organic species on the surface, and this is an interesting fact that needs an explanation. As it has been already proposed for 2005 RR$_{43}$ (Pinilla-Alonso et al.\cite{Pinilla-AlonsoRR43}) different scenarios are possible: the surface has been recently replenished with fresh material (e.g. due to cryovolcanism, multiple collisions or a recent huge collision), or these objects are originally carbon depleted.

An increase of fresh material on the surfaces of icy bodies in the Solar System can be explained by processes of replenishment with fresh and unaltered material from the interior. Subsurface material could be exposed by outgassing onto the surface or by flows from the interior as a result of cryovolcanism. This process would produce a surface composed of patches of fresh water ice covering partially an older surface of processed ices. Cryovolcanism has already been proposed as a mechanism of continuous resurfacing of the surface of medium-size TNOs as (50000) Quaoar (Jewitt \& Luu, \cite{JewittQuaoar}) or Charon (Cook et al., \cite{CookCharon}) but it has always been related to ammonia hydrate. Ammonia hydrate is an expected product of condensation in the high-density protoplanetary disk and evidence from the comets composition show that this molecule is also trapped from the solar nebula (Kawakita \& Watanabe, \cite{KawakitaComets}). For TNOs greater than 400 km in diameter, an ice mixture of ammonia and water would be melted at temperatures typical of the interiors of these bodies. This melted mixture could propagate through cracks until it reaches the surface in a similar way that basaltic volcanism propagates on Earth (Stevenson, \cite{StevensonCryo}).
In this case, the surface would show patches of fresh crystalline water ice with traces of ammonia hydrate covering partially older amorphous water ice. In the spectrum, we should see signs of ammonia hydrate clearly evidenced by an absorption band at 2.2$\mu$m. This band is  not present in the spectrum of 2003 EL$_{61}$. Also our fits favour an intimate mixture instead of an areal one, so cryovolcanism is not a scenario compatible with the observations. Furthermore, as the surface properties of the group of TNOs related to 2003 EL$_{61}$ are similar, we should conclude that resurfacing mechanism acting on them is similar. But as several of them have diameters smaller than 400 km, we can discard cryovolcanism.

Another scenario is that of resurfacing produced by collisions. One collisional event has been proposed by Brown et al. (\cite{Brownfamily}) as the origin of the group of TNOs with similar orbital and spectral properties to those of 2003 EL$_{61}$'s. Such an energetic impact would release a large amount of energy that could partially be converted into heat and that would produce a significant increase of the temperature (above 100-130 K) leading to the crystallization of a large ammount of water ice. The eroded material would be sublimated and globally distributed over the TNO surface on a timescale of tens of hours (Stern \cite{Stern2002}). The result of this process would be a fresh surface covered with crystalline water ice as proposed for 2002 TX$_{300}$ (Licandro et al. \cite{LicTX}), another member of the group. After the event the temperature of the ice cools down to about 40 K and irradiation process starts to transform crystalline into amorphous water ice.
From the analysis of the spectrum of 2003 EL$_{61}$ and the models presented in Sec 4., we find a percentage of crystalline and amorphous water ice on the surface in a ratio of 1:1, and this can be used to constrain the age the surface of 2003 EL$_{61}$.
Leto \& Baratta \cite{Leto03} found from laboratory experiments that a 1:1 mixture of crystalline and amorphous water ice is obtained by irradiating crystaline water ice with a dose of at least $1eV/molecule$. Taking also into account the percentage of crystalline and amorphous ice in our model and the results of Strazzulla et al. (\cite{Strazzulla03}) about the time needed to produce a radyolitic change in irradiated water ice, we can infer a timescale for this process of about $10^8\; yr$ at an heliocentric distance of $40\; AU$ (the error in this estimate is of a factor of $5$). Moreover, we have to consider the effect of collisions with smaller particles in the TNb, these continuous events recrystallize water ice on the surface and compete with that of amorphization by irradiation. So, the timescale needed to produce a 1:1 crystalline/amorphous surface is probably longer than $10^8\; yr$.

This estimation puts a constraint on the age of the surface of $2003\; EL_{61}$, since a recent ($<10^8\; yr$) impact that rejuvenated and recrystallized the whole surface has to be excluded. It is important to note that this result agrees with recent calculations indicating that the large impact that most probably originated the family of $2003\; EL_{61}$ should be primordial (Ragozzine \& Brown, \cite{Ragozzine}).

In any case, the surface of 2003 EL$_{61}$ should be older than $10^8\; yr$, i.e it is old enough to show on its surface the signal of complex organics, but this is not the case for this object. The only reasonable explanation for the lack of organics on the surface of the members of this population is that they have a composition depleted of carbon chains (as already suggested for 2005 RR$_{43}$ (Pinilla-Alonso et al.\cite{Pinilla-AlonsoRR43})). This carbon depleted population turns into the first group whose surface properties can only be explained by a different original composition, and this needs to be further investigated. Either the depletion of carbon chains is related to the hypothetic giant collision or alternatively these TNOs were formed in an heterogeneous region of the solar nebula.

All the members of this group have similar orbital parameters (41.6 $<$a$<$ 43.6 AU, 25.8 $<$i$<$ 28.2 deg., 0.10 $<$e$<$ 0.19) and lie in an unstable region of the TNb crossed by Neptune resonances. This suggests some of these objects could be injected to inner regions of the TNb finishing their lives as Jupiter family comets. Consequently this group could be a source (not necessarily unique) of the carbon depleted comets in the Jupiter family already noticed by A'Hearn et al. (\cite{AHearn95}).

\section{Conclusions}

We observed 2003 EL$_{61}$ on 2005 August 1.92 UT simultaneously in the visible and near infrared with two telescopes, the 4.2m WHT and the 3.6m TNG at the ORM (Canary Islands, Spain); and only in the near-infrared with the TNG on 2006 February 25, from 01:21 to 04:33 UT. We present the low-resolution spectra obtained in both nights. The similarities between near infrared spectra obtained at different rotational phases (covering almost 80 \% of the rotational period) let us conclude that the surface of this TNO is almost homogeneous.

Fitting 2003 EL$_{61}$  spectrum using Hapke (\cite{HapkeModels}, \cite{Hapke93}) models we concluded that it is composed of almost pure water ice. We reproduce the spectrum with an intimate mixture of crystalline and amorphous water ice in a proportion 1:1. We also obtain a good fit with a multi-layer model representing a thin cap of an intimate mixture of crystalline and amorphous water ice in a proportion of 4:6 over an infinite layer of crystalline water ice. Both configurations are physically consistent with the homogeneity of the surface and with the processing of water ice under cosmic irradiation.

If the present surface of 2003 EL$_{61}$ is the product of a large collision, the relative percentage of amorphous and crystalline ice we obtain from the models, implies that this collision should have happened more than 10$^{8}$ years ago, in agreement with dynamical models.

Furthermore, these models evidence that adding other constituents in a percentage over 8\% impoverishes the fit of the spectrum introducing new features or reddening the slope in the visible rage. So we discard the presence of a significant amount of other components on the surface but water to a high level of confidence. In particular we discard the presence of complex organics.

Considering that our models favor an intimate 1:1 mixture of amorphous/crystalline water ice, and that there is no signature of ammonia hydrate in the spectrum, we conclude that cryovolcanism is unlikely the main resurfacing mechanisms on 2003 EL$_{61}$.

Given the estimation of the age and the lack of an evolved mantle of complex organics on the surface of 2003 EL$_{61}$, we conclude that, as for 2005 RR$_{43}$, and probably the other members of the EL$_{61}$ group of TNOs identified by Pinilla et al.  \cite{Pinilla-AlonsoRR43}, this body has a significant smaller fraction of carbon chains on its surface. This group is a carbon-depleted population of TNOs that, considering their orbital parameters, could be the origin of some of the carbon-depleted comets already noticed by A'Hearn et al. (\cite{AHearn95}).

\begin{acknowledgements}
Based on observations made with the Italian Telescopio Nazionale Galileo (TNG) operated on the island of La Palma by the Fundaci\'on Galileo Galilei of the INAF (Istituto Nazionale di Astrofisica) at the Spanish Observatorio del Roque de los Muchachos of the Instituto de Astrof\'isica de Canarias. R.B and G.S are greatful to Italian Space Agency contract n. 1/015/07/0. T.L.R acknowledges support from NASA's Planetary Geology and Geophysics program. We acknowledge F. Merlin and S.C. Tegler for sending us their spectra of 2003 EL$_{61}$.

\end{acknowledgements}


\begin{thebibliography}{}
\bibitem[1995]{AHearn95}
A'Hearn, M. F., Millis, R. L., Schleicher, D. G. et al., 1995, Icarus,  118, 223.
\bibitem[1991]{Baratta91}
Baratta, G.A.,Leto, G., Spinella, F. et al. 1991, A\&A, 252, 421.
\bibitem[1998]{BrownHA}
Brown, R.H., Cruikshank, D.P., Tokunaga, A.T. et al. 1988, Icarus, 74, 262.
\bibitem[2007]{Brownfamily}
Brown, M.E., Barkume, K.M., Ragozinne, D. et al., 2007, Nature, 446, 294.
\bibitem[2006]{Brunetto-Colors} 
Brunetto, R., Barucci,M.~A., Dotto, E., \& Strazzulla, G.\ 2006, \apj, 644, 646 
\bibitem[2007]{Brunetto-Pyroxenes}
Brunetto,R., Roush, T.L., Marra, A.C., Orofino, V., 2007, Icarus, 191, 381.
\bibitem[2008]{Brunetto-Roush}
Brunetto,R. \& Roush, T.L., 2008, A\&A (accepted).
\bibitem[1993]{Clark}
Clark, R.N., G.A. Swayze, A.J. Gallagher et al., 1993, USGS Open File Report 93-592, 1340.
\bibitem[2007]{CookCharon}
Cook, Jason C., Desch, Steven J., Roush, Ted L. et al., 2007, ApJ, 663, 1406.
\bibitem[1991]{Cruikshank}
Cruikshank, D. P., Allamandola, L. J., Hartmann, W. K. et al., 1991, Icarus, 94,345.
\bibitem[2004]{Fornetal2004}
Fornasier, S., Dotto, E., Barucci,A. et al, 2004, A\&A 422, 43.
\bibitem[2002]{Gil-Hutton}
Gil-Hutton, R., 2002, P\&SS 50, 57.
\bibitem[1998]{Grundywater}
Grundy, W. M and Schmidtt, B. 1998 JGR, 103, 25809.
\bibitem[2002]{GrundyMet}
Grundy, W. M.; Schmitt, B. \& Quirico, E., 2002, Icarus, 155, 486.
\bibitem[1981]{HapkeModels}
Hapke, B., 1981, J. Geophys. Res., 86, 3039
\bibitem[1993]{Hapke93}
Hapke, B.\ 1993, Theory of Reflectance and Emittance Spectroscopy (New York: Cambridge Univ. Press)
\bibitem[2004]{JewittQuaoar}
Jewitt,D.C. \& Luu, J., 2004, Nature, 432, 731
\bibitem[1984]{Johnetal84}
Johnson R., Lanzerotti L., \& Brown W., 1984, Adv. Space Res. 4, 41.
\bibitem[1994]{Khare}
Khare,B.N., Sagan,C., Thompson, W.R. et al., 1994, Can.J.Chem., 72, 678.
\bibitem[2002]{KawakitaComets}
Kawakita, H \& Watanabe, J. 2002, ApJ. 572, L177.
\bibitem[1992]{Landolt}
Landolt, A., 1992, AJ 104, 340.
\bibitem[2003]{Leto03}
Leto, G., \& Baratta,G.~A.\ 2003, \aap, 397, 7
\bibitem[2008]{LacerdaEL61}
Lacerda, P., Jewitt, D., \& Peixinho, N., 2008, AJ Accepted.
\bibitem[2001]{Licreduc}
Licandro, J., Oliva E., \& Di Martino M., 2001, A\&A 373, L29.
\bibitem[2006]{LicTX}
Licandro, J., di Fabrizio, L., Pinilla-ALonso, N. et al. 2006, A\&A 457, 323.
\bibitem[1992]{Massey1992}
Massey, P., Valdes, F. \& Barnes, J., 1992, in "A User's Guide to Reducing Slit
Spectra with IRAF", http://iraf.noao.edu/iraf/ftp/iraf/docs/spect.ps.Z.
\bibitem[2007]{MerEL61}	
Merlin, F., Guilbert, A., Dumas, C. et al., 2007, A\&A 466,  1185.
\bibitem[1983]{Mooreetal83}
Moore, M., Donn, B., Khanna, R, \& A'Hearn, M., 1983, Icarus 54, 388.
\bibitem[2007]{Pinilla-AlonsoRR43}
Pinilla-Alonso, N., Brunetto, R., Licandro,J. \& Gil-Hutton,R., 2007, A\&A 468, L25.
\bibitem[2006]{RabinEL61}
Rabinowitz, D. L., Barkume, K., Brown, M. E. et al., 2006, ApJ,639, 1238
\bibitem[2007]{Ragozzine}
Ragozzine, D. \& Brown, M.E., 2007, AJ, 134,2160.
\bibitem[1998]{Schmitt}
Schmitt, B., Quirico, E., Trotta, F. \& Grundy, W.M., 1998, Solar System Ices (Kluwer Academic Publisher)
\bibitem[1980]{Sill}
Sill, G. T.; Fink, U.; Ferraro, J. R., 1980, JOSA, 70, 724.
\bibitem[2004]{StevensonCryo}
Stevenson, D.J., 2004, Nature, 432, 681
\bibitem[2002]{Stern2002}
Stern, A., 2002, AJ 124, 2297.
\bibitem[1991]{StrazzullaJo91}
Strazzulla G., \& Johnson R., 1991, in {\it Comets in the Post-Halley era}, R.L. Newburn Jr., M. Neugebauer and J. Rahe, eds., Kluwer Academic Publishers, Netherlands, 243.
\bibitem[2003]{Strazzulla03}
Strazzulla,G., Cooper,J.F., Christian, E.R., \& Johnson,R.E., 2003, C.R. Phys., 4, 791.
\bibitem[2007]{TegEL61}
Tegler, S. C., Grundy, W. M., Romanishin, W. et al., 2007, A\&A 133, 526.
\bibitem[2007]{TrujEL61}	
Trujillo, C. A., Brown, M., Barkume, K. M. et al., 2007 ApJ 655, 1172.
\bibitem[1984]{Warren84}
Warren, S. G., 1984 ApOpt, 23, 1206-1225
\end{thebibliography}
\end{document}